\documentclass[conference]{IEEEtran}
\IEEEoverridecommandlockouts
\usepackage{cite}
\usepackage{amsmath,amssymb,amsfonts,amsbsy}
\usepackage{algorithm}
\usepackage[noend]{algpseudocode}
\usepackage{graphicx}
\usepackage{textcomp}
\usepackage[table]{xcolor}    % loads also »colortbl«
\usepackage{cases}
\usepackage{stfloats}

\usepackage[letterspace=-125]{microtype}

\usepackage{booktabs}
\usepackage{adjustbox}
\usepackage{multirow}
\usepackage{enumitem}
\usepackage{titlesec}

\usepackage{stackengine}

\newcommand\blfootnote[1]{
  \begingroup
  \renewcommand\thefootnote{}\footnote{#1}
  \addtocounter{footnote}{-1}
  \endgroup
}

\titlespacing{\subsection}{1em}{\parskip}{-\parskip}
\titlespacing{\subsubsection}{1em}{\parskip}{-\parskip}

\newcommand{\cf}[1]{$\mathtt{#1}$}
\newcommand{\cfm}[1]{\mathtt{#1}}

\newcommand{\mE}{\mathcal{E}}

\newcommand{\mG}{\mathcal{G}}

\newcommand{\mI}{\mathcal{I}}

\newcommand{\mN}{\mathcal{N}}
\newcommand{\mO}{\mathcal{O}}

\newcommand{\mV}{\mathcal{V}}

\newcommand{\hem}{\hspace{1em}}

\newcommand\mydots{\ifmmode\ldots\else\makebox[1em][c]{.\hfil.\hfil.}\thinspace\fi}

\newcommand{\mcl}{\multicolumn}

\newcommand{\binzero}{\mathtt{0}}
\newcommand{\binone}{\mathtt{1}}

\def\BibTeX{{\rm B\kern-.05em{\sc i\kern-.025em b}\kern-.08em
    T\kern-.1667em\lower.7ex\hbox{E}\kern-.125emX}}
\begin{document}

% \title{Towards Multiphase Clocking in Single-Flux Quantum Systems
% \vspace{-10pt}
% }

\title{\adjustbox{max width=\textwidth}{Towards Multiphase Clocking in Single-Flux Quantum Systems}\vspace{-10pt}
}

\author{
\IEEEauthorblockN{Rassul~Bairamkulov and Giovanni~De~Micheli}
\IEEEauthorblockA{Integrated Systems Laboratory, EPFL}
Lausanne, Switzerland \\
rassul.bairamkulov@epfl.ch, giovanni.demicheli@epfl.ch
\vspace{-15pt}
}

\maketitle

\begin{abstract}

Rapid single-flux quantum (RSFQ) is one of the most advanced superconductive electronics technologies. 
SFQ systems operate at tens of gigahertz with up to three orders of magnitude smaller power as compared to CMOS. 
In conventional SFQ systems, most gates require clock signal. 
Each gate should have the fanins with equal logic depth, necessitating insertion of path-balancing (PB) DFFs, incurring prohibitive area penalty. 
 
Multiphase clocking is the effective method for reducing the path-balancing overhead at the cost of reduced throughput. 
However, existing tools are not directly applicable for technology mapping of multiphase systems. 
To overcome this limitation, in this work, we propose a technology mapping tool for multiphase systems. 
Our contribution is threefold. 
First, we formulate a phase assignment as a Constraint Programming with Satisfiability (CP-SAT) problem, to determine the phase of each element within the network.
Second, we formulate the path balancing problem as a CP-SAT to optimize the number of DFFs within an asynchronous datapath. 
% Second, we propose a method to identify the independent datapaths within the network for more efficient processing. 
Finally, we integrate these methods into a technology mapping flow to convert a logic network into a multiphase SFQ circuit. 
In our case studies, by using seven phases, the size of the circuit (expressed as the number of Josephson junctions) is reduced, on average, by 59.94 \% as compared to the dual (fast-slow) clocking method, while outperforming the state-of-the-art single-phase SFQ mapping tools.
\end{abstract}

% \begin{IEEEkeywords}
% component, formatting, style, styling, insert
% \end{IEEEkeywords}

\blfootnote{This work was supported by the SNF grant ``Supercool: Design methods and tools for superconducting electronics'' under Grant 200021-1920981.

The authors thank Professor Peter Beerel from the University of Southern California for the insightful discussions.}

\begin{figure*}[b!]
\centering
\vspace{-1em}
\hrule
\vspace{0.2em}
\stackunder{\includegraphics[width=0.32\textwidth]{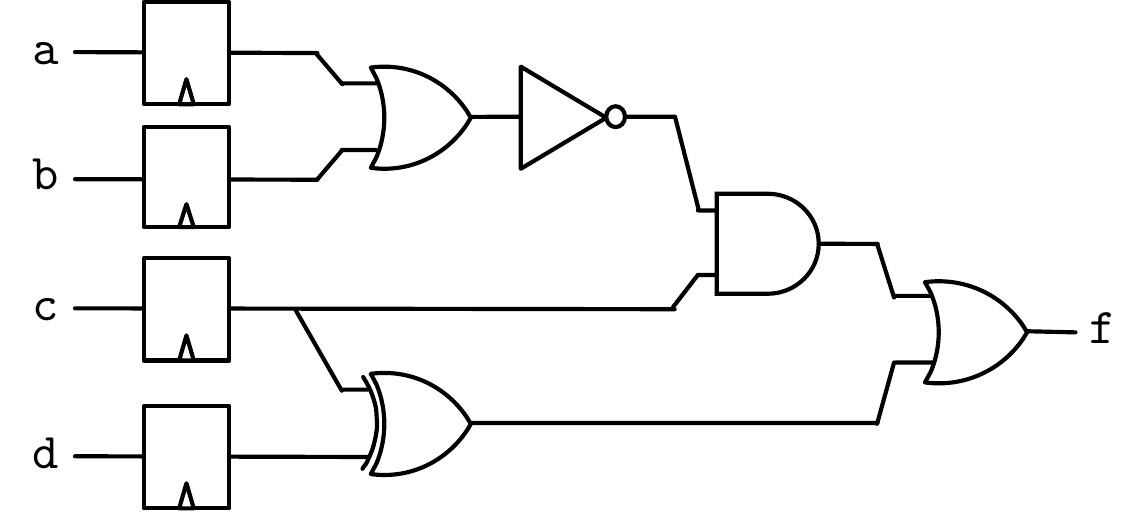}}{a)} \hfill
\stackunder{\includegraphics[width=0.32\textwidth]{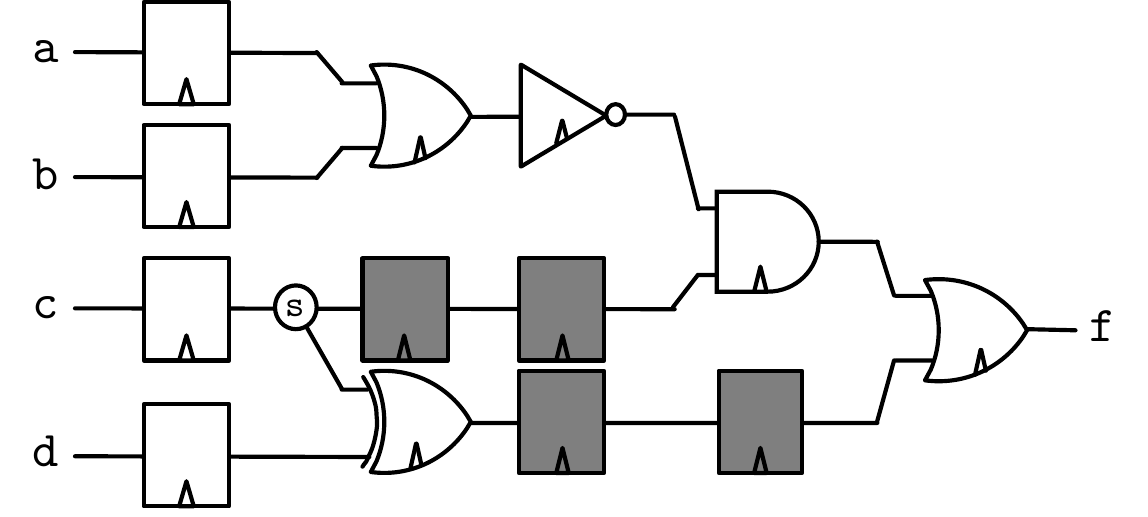}}{b)}
\hfill
\stackunder{\includegraphics[width=0.32\textwidth]{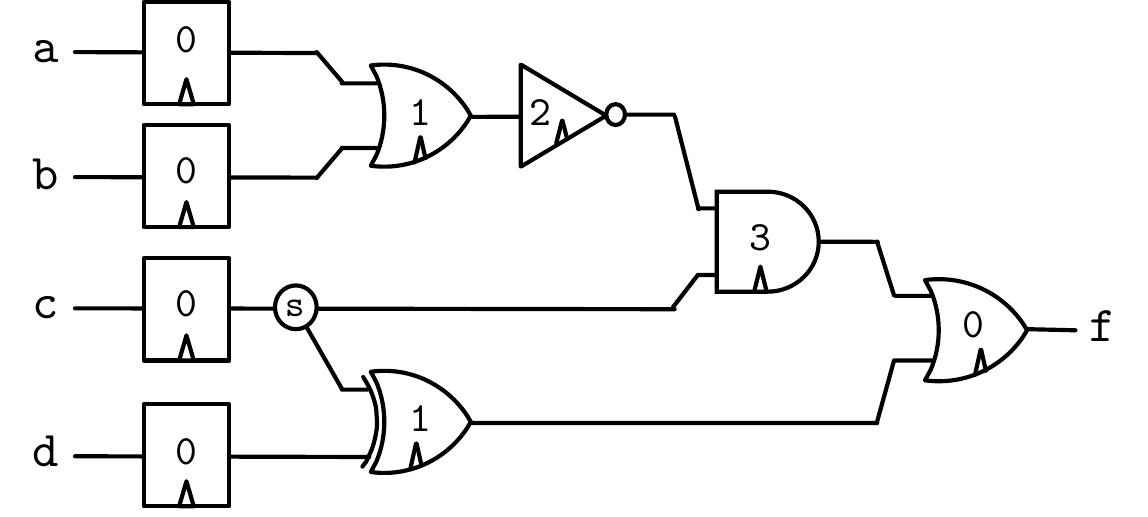}}{c)}
\caption{
% Technological constraints in RSFQ. 
a) An example of a CMOS circuit.
b) Equivalent SFQ circuit with a splitter and two path balancing DFFs.
c) A four-phase SFQ realization with no path balancing DFFs.
The numbers indicate the phase corresponding to each clocked gate.
}
\vspace{-1.4em}
\label{fig:spl_pb}
\end{figure*}

\section{Introduction}

Superconductive electronics is one of the most promising beyond-CMOS technologies, with Rapid Single-Flux Quantum (RSFQ)~\cite{likharev1985} as the most prominent superconductive technology.
RSFQ systems consistently achieve operating frequencies on the order of tens of gigahertz~\cite{kawaguchi2015,deng1997data}, with particular structures accelerating to hundreds of gigahertz~\cite{chen770ghz,herr2002high}. %jabbari2020repeater
Furthermore, the operating power of the RSFQ systems is two to three orders of magnitude smaller than CMOS, even considering the refrigeration power~\cite{holmes2013energy}. 
Due to the low power, low noise, and high speed RSFQ systems are utilized in high-resolution sensors \cite{sternickel2006biomagnetism} and high-performance wireless communication systems \cite{tuttlebee2002software}.
% Increasing the integration density of the 
In addition, these advantages make the RSFQ a promising candidate for large-scale stationary computing systems (e.g., data centers)~\cite{holmes2013energy}, as well as low-power computing in space~\cite{krylov2022book}. 

Elevating complexity of RSFQ systems is however a challenging task.
Due to the major differences between RSFQ and CMOS technologies, standard design methodologies and tools are not applicable to RSFQ.
% Realizing large-scale computing systems remains a major challenge in RSFQ due to its reliance. 
In CMOS, transistors encode information by voltage levels. 
In contrast, in RSFQ, Josephson Junctions (JJ)  encode information using single flux quantum (SFQ) pulses~\cite{bunyk2001}.
\footnote{
% In-depth description on RSFQ technology can be found in [8] and [9] for those readers who are particularly interested
For an in-depth discussion of the SFQ technology we refer an interested reader to \cite{krylov2022book, bunyk2001}.}
The absence or presence of a SFQ pulse encode, respectively, a logic $\binzero$ or $\binone$. 
The flow of pulses within a logic network requires synchronization. 
While a few structures, such as splitter and \cf{OR}, can operate asynchronously, most logic gates in the original RSFQ cell library, such as \cf{AND}, \cf{NOT} and \cf{XOR}, require clock signal~\cite{likharev1985}.
This feature requires the data to be pipelined at the gate level.

Two features further complicate scaling the RSFQ systems. 
Due to the quantized nature of SFQ pulses, most RSFQ logic gates are limited to fanout of one. 
A special gate called \textit{splitter} is necessary to duplicate a signal~\cite{krylov2022book, bunyk2001}, as illustrated in Fig.~\ref{fig:spl_pb}b. 
% The need for splitters is particularly significant in clock distribution networks.
% Delivering a clock signal to $n$ clocked gates typically requires a binary tree of $(n-1)$ splitters~\cite{self_qucts}.
% This inherent feature enables a significant increase in horizontal parallelism by increasing the logical pipeline depth~\cite{self_qucts}. 
In addition, path balancing is required to ensure the correct order of data propagation within the network, as indicated by four path balancing D-flip-flops (DFF) in Fig.~\ref{fig:spl_pb}b.
The number of path balancing DFFs and splitters can be prohibitively large, degrading the area and yield.
The complexity of the clock tree synthesis is also increased due to the large number of clocked elements. 
% Path-balancing-aware technology mapping is proposed in~\cite{pbmap2019}.
% Using dynamic programming, the number of path balancing DFFs is reduced, on average, by 21\%, yielding a 12\% smaller area. 
Despite the advances in technology mapping~\cite{pbmap2019, kito2021logic}, the overhead of path balancing and clock distribution  remains large, motivating the investigation of alternative RSFQ architectures. 

% Two approaches to alleviating the path balancing overhead are proposed in the literature. 
% The first approach targets the minimization of the path balancing overhead by expanding the functionality of SFQ logic achievable in a fewer clock cycles. 
% Gate compounding technique~\cite{self_glsvlsi} is a novel technique for maximizing the depth of an SFQ circuit by maximizing the use of the asynchronous elements. 
% Additional logic elements, such as \mathtt{XNOR} and \mathtt{NIMPLY}, are realized as single-cycle gates.
% In Fig.~\ref{fig:spl_pb}d, for example, a compound gate $\neg(\mathtt{a}+\mathtt{b})\oplus\mathtt{c}$ is used to reduce the pipeline depth and eliminate the PB DFFs.

% The second approach explores alternative clocking methods to relax the requirement of path balancing. 

Alternative clocking methods have been explored in the literature to alleviate the issue of path balancing. 
In dual clock methodology~\cite{dual_clock}, a logic circuit is partitioned into separate clocking domains.
% using the architectural nondestructive readout (NDRO) registers operating at low frequency. 
The gates within a single domain are clocked at high frequency $f_{hi}$, while cross-domain communication is performed via the nondestructive readout (NDRO) registers operating at the low frequency \mbox{$f_{lo}=f_{hi}/N$}. 
In~\cite{dual_clock}, the number of path balancing DFFs is reduced by 35\% with \mbox{$N=5$}~\cite{dual_clock}. 
The throughput of the system is however reduced by a factor of 5, establishing a tradeoff between the area and the throughtput of the system.
The primary limitation of the dual clock method is the use of the large registers to repeat the data signals and additional \cf{AND} gates to suppress the spurious pulses within the system~\cite{pasandi_acm}.
In addition, similar to full path balancing, distributing the high frequency clock signal makes the timing closure more challenging to achieve.  
% These extra gates complicate the design and reduce the area savings brought by this approach. 

% The number of DFFs can be further reduced by using greater ratio.
% This separation relaxes the requirement for full path balancing at the cost of reduced throughput. 

Multiphase clocking has recently been proposed for designing area-efficient SFQ systems~\cite{li_beerel}. 
% The circuit operates u
Using several phase-shifted clock signals allows data to propagate to the subsequent stages without DFFs. 
For example, by using four-phase clocking, the path balancing DFFs can be completely removed, as shown in Fig.~\ref{fig:spl_pb}c.  
% to achieve path balancing with fewer DFFs. 
Using multiple phases was demonstrated effective in reducing the SFQ circuit area, thanks to the reduction in DFF count (i.e., smaller logic network) and  fewer clocked elements yielding smaller clock network size.
With only two phases, the number of DFFs and area are reduced by, respectively, 55\% and 41\%~\cite{li_beerel}, despite the overhead of generating and distributing separate clock signals.
% Further reduction brought by a
Additional phases further reduce the system area. 
% Adding  more phases further reduces the path balancing overhead.
Furthermore, in an $n$-phase system, the clock signals have $n$ times lower frequency, simplifying the clock distribution process.
Similar to dual clock method, the area reduction is achieved by sacrificing the throughput.
An $n$-phase system has $n$ times smaller throughput as compared to a single phase system. 
% In addition 

% The technique was demonstrated to significantly reduce the path balancing overhead in SFQ circuits. 

% The primary limitation of this technique is the creation of additional clock distribution network for each clock phase.
% Note, however, that while PB DFFs occupy a relatively valuable device layers, the clock network occupies the less critical metal layers, justifying this tradeoff for small number of phases. 
% % Nevertheless, the benefit of reducing PB overhead should therefore outweigh the cost of an extra clock distribution network.
% Another limitation of multiphase clocking is the complex design of clock distribution networks. 
% Each clocked element should be assigned to a particular phase to reducing the number of PB DFFs while ensuring correct functionality. 
% ILP-based methodology for phase assignment is presented in the seminal work of~\cite{beerel_multiphase}. 
% This method however scales poorly with the circuit size due to the large number of conditional variables. 
% Furthermore, the cost function does not support the recovergent paths with shared DFFs. 

% Although gate compounding and multiphase clocking utilize different dimensions to tackle path balancing, no methodology exists in the literature to combine these two techniques. 

Although the multiphase clocking has been shown to reduce the path balancing overhead, existing technology mapping tools offer limited support for the multiphase systems. 
For example, path balancing DFFs cannot be efficiently distributed within the complex datapaths utilizing asynchronous SFQ gates. 
% The asynchronous elements within the compound logic gates render direct application of multiphase clocking to compound gate circuits suboptimal. 
To bridge this gap, in this paper, we propose a novel methodology to apply the multiphase clocking to SFQ circuits. 
Our contribution is threefold,
% The proposed technique allows combining the gate compounding technique with multiphase clocking, significantly outperforming 
% reducing the number of PD DFFs with fewer clock phases. 
\begin{enumerate}[leftmargin=*]
\item we extend the DFF insertion methodology to assign the phase to each gate within the circuit, including unclocked elements.
% a clock stage and phase is assigned to each gate using the ILP-based algorithm. 
% First, independent asynchronous datapaths are identified within the circuit. 
\item we formulate PB DFF insertion as a Constraint Programming with SAT (CP-SAT) problem to satisfy the timing constraints with minimum number of PB DFFs. 
We improve the scalability by identifying independent paths allowing the asynchronous paths to be processed separately.
\item we integrate phase assignment and DFF insertion into the technology mapping flow to realize an arbitrary logic network  with multiple clock phases. 
\end{enumerate}
With only two phases, our flow outperforms the state-of-the-art single-phase methods.
Compared with the dual clocking method, our technology mapping flow achieves up to 82\% smaller size, while offering equal throughput.
% using compound-gates
% The proposed methodology demonstrates the significant potential of MPC. 
% isolated application of gate compounding or multiphase clocking.
% In our case studies, the number of path balancing DFFs is reduced, on average, by \AVGdffCGMP\%, as compared to the multiphase clocking with conventional SFQ gates, and \AVGareaCGMP\% as compared to single-phase gate compounding.

The rest of the paper is organized as follows.
The principles of SFQ technology and multiphase clocking are reviewed in \mbox{Section~\ref{sec:background}}.
The CP-SAT-based method for phase assignment is presented in \mbox{Section~\ref{sec:phase_assignment}}.
PB DFF insertion are described in \mbox{Section~\ref{sec:dff_insertion}}.
Experimental results and comparison with the prior works are provided in \mbox{Section~\ref{sec:validation}}, followed by conclusions in \mbox{Section~\ref{sec:conclusions}}.

\section{Background} \label{sec:background}

A logic network $\mN=(\mV,\mE)$ is a directed acyclic graph (DAG) where $\mV$ is a set of nodes and $\mE\subseteq \mV\times \mV$ is a set of edges. 
The set of nodes $\mV=\mI\cup \mO\cup \mG$ consists of three disjoint subsets representing, respectively, the primary inputs (PI), primary outputs (PO) and gates within the network.
A set of fanins $FI(g)$ (fanouts $FO(g)$) of gate $g\in \mV$ denotes the nodes connected to $g$ via and incoming (outgoing) edge. 
% consider defining mathmematically
% The PIs (POs) are the nodes with no incoming (outgoing) edges.

Technology mapping refers to the process of transforming an arbitrary logic network $\mN$ into a technology-specific representation $\mN'$ using a particular set of primitives.
The technology mapping aims to optimize the target characteristics of the mapped network $\mN'$, such as area, delay, or yield
% The target technology defines the type and properties of primitives, typically using the standard cells. 
% The technology mapping aims to optimize the target characteristics of the mapped network $\mathcal{N}'$, such as area, delay, or yield.
The nature of a technology greatly influences the technology mapping process.
In the next subsection, the properties of the SFQ technology are reviewed. 

\subsection{Single-Flux Quantum technology}

Rapid Single-flux Quantum (RSFQ) is a cryogenic superconductive computing logic family based on Josephson junctions (JJ). 
% The number of JJs is highly correlated with the total area of a SFQ circuit and is therefore commonly used as an area metric \cite{krylov2022book}.
RSFQ gates consist of superconductive loops storing quantized magnetic flux. 
The information is transferred between the superconductive loops in form of SFQ pulses with the area of $\Phi_0=\hbar/2e\approx 2.07$ mV$\cdot$ps~\cite{krylov2022book}.
% The SFQ logic gates are typically clocked, where a logic zero (one) is denoted by the absence (presence) of a SFQ pulse within a clock period.
Based on synchronization mechanisms, the SFQ logic gates can be divided into three major categories~\cite{bairamkulov_glsvlsi23}.
\begin{itemize}[leftmargin=*]
\item \textbf{Asynchronous input, Asynchronous output (\cf{AA})} components process the input information immediately upon arrival. 
The most common gates in this category are splitter and merger.
A \textit{splitter} produces two pulses for each incoming pulse. 
A \textit{merger} directs signals from two input branches into one output branch, effectively performing an \cf{OR} function.
% Note that the merger produces two subsequent output pulses if input pulses are temporally separated, or a single pulse, if input signals arrive simultaneously.
\item \textbf{Asynchronous input, Synchronous output~(\cf{AS})} elements process the input information immediately upon arrival and release the output synchronously after the arrival of the clock signal.
The components of this type include \textit{D-flip-flop~(DFF)}, \textit{inverter~(\cf{NOT})} and \textit{exclusive-or~(\cf{XOR})}.
% , illustrated in Figs.~\ref{fig:AS_elems}. 
\item \textbf{Synchronous input, Asynchronous output~(\cf{SA})} elements require the input signals to arrive simultaneously to operate correctly. 
The result of the computation is released immediately after processing.
Assuming inputs arrive simultaneously, a CB can be tuned to produce at most a single output pulse, producing an \textit{\cf{OR}} element~\cite{mukhanov1987ultimate}. 
% depicted in Fig.~\ref{fig:splitter}c 
Furthermore, by adjusting the JJ size and bias current, the \cf{OR} structure can be transformed into \cf{AND} element, requiring both signals to arrive simultaneously to produce an output pulse.
\end{itemize}
In conventional RSFQ~\cite{likharev1985}, the \cf{OR} and \cf{AND} gates incorporate the DFFs at the inputs to ensure the simultaneous release of the data pulses.
However, the simultaneity can be achieved by other \cf{AS} gates, such as \cf{NOT} and \cf{XOR}~\cite{bairamkulov_glsvlsi23}, allowing richer functionality to be realized in fewer clock cycles.

To correctly consider the type of each element, we divide the set of logic gates into three disjoint subsets \mbox{$\mG=\mG_{\cfm{AA}}\cup \mG_{\cfm{AS}}\cup \mG_{\cfm{SA}}$}, where each subset represents the elements of the corresponding category. 
Note that only gates in the subset $\mG_{\cfm{AS}}$ require clock signal.

% Advanced mapping techniques developed for CMOS technology are not directly applicable to SFQ technology, due to the fundamental differences between the technologies. 
% The goal of the technology mapping process is to transform the \textit{technology-independent} representation of a logic network, such as And-Inverter-Graph (AIG), into a \textit{technology-dependent} representation $\mathcal{N}'=(\mV',\mE')$.
% The mapping procedure attempts to reduce area, delay, or a combination of area and delay in the final LUT network.

\subsection{Multiphase clocking}

% Conventional synchronous networks are based on single-phase clock signal, where all clocked elements receive the same clock signal. 
% % The data flow within these systems requires balancing. 
% As shown in Fig.~\ref{fig:spl_pb}b, to balance the paths differing by $n$ cycles, $n$ DFFs are needed.
% This issue is less critical in CMOS where an arbitrary number of logic elements can be inserted within a single clock stage, provided setup and hold time constraints are satisfied. 
% In RSFQ, however, most functional gates are clocked limiting the functionality achievable within a single clock period. 
% Path balancing is therefore a major issue in SFQ technology, necessitating placement of additional DFFs to ensure equal logic level within the network.

\begin{figure}[b]
\centering
\includegraphics[width=0.9\columnwidth]{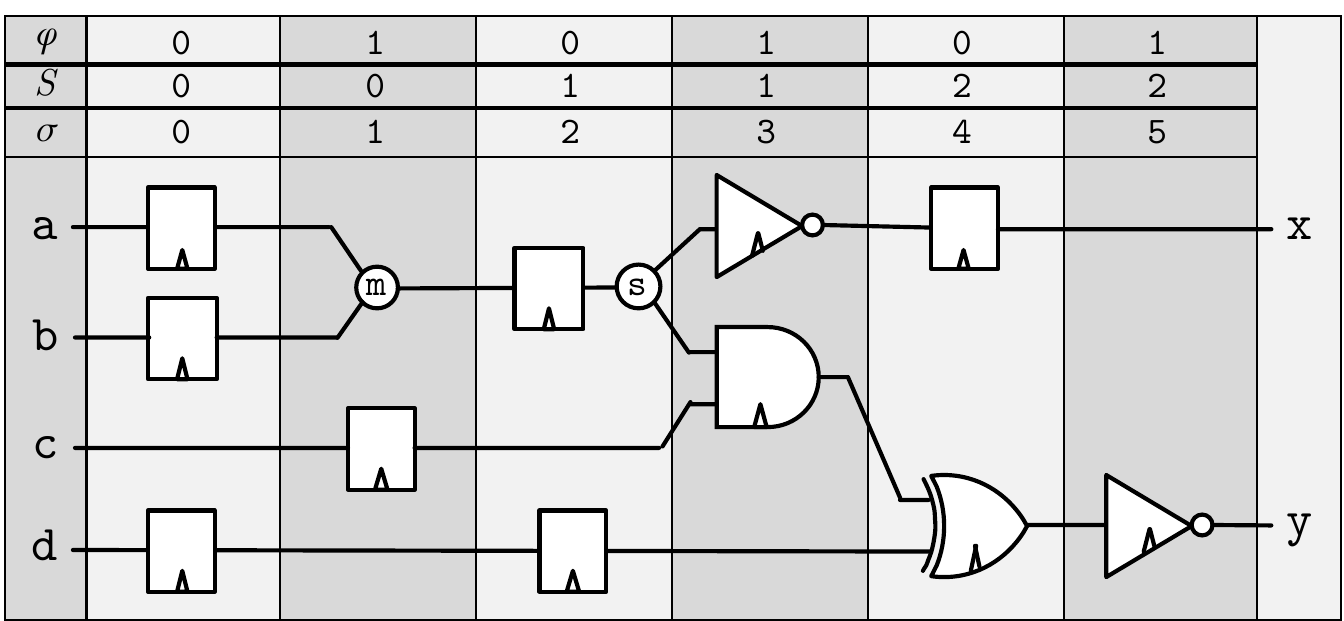}
% timing_defs
\caption{
An example two-phase network.
Each PI is placed at stage 0 or 1, at epoch 0.
The POs are placed at stages 4 and 5, corresponding to epoch 2.
}
\label{fig:timing}
\end{figure}

Multiphase clocking (MPC) is a clocking technique based on several clock signals with equal clock period.
The seminal work by Li~\textit{et~al.}~\cite{li_beerel} demonstrated MPC as an effective remedy for path balancing in SFQ systems.
% In this subsection, we outline the fundamental principles of multiphase clocking first outlined in~\cite{sakallah}. 
A $n$-phase system utilizes $n$ periodic signals $\{t_0,\cdots,t_{n-1}\}$ operating at the same frequency and different phases. 
%$f$
% , $0\leq \varphi_0 < \cdots < \varphi_{n-1} \leq 2\pi$.
Each clocked element $g\in \mG_{\cfm{AS}}$ within the network is synchronized by only one clock signal at phase $\varphi(g)$. 
The epoch $S(g)$ of a gate $g$ is defined as the number of clock cycles separating the gate $g$ from the PIs, as illustrated in Fig.~\ref{fig:timing}. 
The clock signals are ordered by phase $\varphi\in\{0,\cdots,n-1\}$, i.e., during any epoch, the clock signal $t_i$ arrives before clock signal $t_j$ if $i<j$.
For consistency in I/O timing, the PIs are placed at the same epoch $0$ and the POs are placed at the same epoch.
For example, the POs are  Fig.~\ref{fig:timing}, the fanouts $\mathtt{x}$ and $\mathtt{y}$ are both placed at epoch 2.
Observe that a path balancing DFF is inserted before $\mathtt{x}$ to equalize the epochs of both POs.
For convenience, we define a \emph{stage} $\sigma(g)$ of a gate $g$ as 
\begin{equation}
    \sigma(g)=nS(g)+\varphi(g).
\end{equation} 

Multiphase clocking relaxes the path balancing requirements of a clocked system.
With $n$ phases, the stage difference $\Delta\sigma(i,j)$ between any pair of sequentially adjacent elements $(i,j)$ should be no greater than $n$.
% \begin{equation}
%     \sigma(j) - \sigma(i) \leq n.
% \end{equation}
Compared with single-phase systems where a $q$-stage difference is balanced by $q$ DFFs, the same difference within an $n$-phase system can be balanced with only $\left\lfloor \frac{q}{n} \right\rfloor$ DFFs, greatly reducing the path balancing overhead in imbalanced paths.

To realize the potential of multiphase clocking in SFQ systems, we propose the two-step DFF insertion methodology based on CP-SAT.
During the phase assignment step, the SFQ gates are assigned a phase to reduce the number of DFFs within the system, as described in \mbox{Section~\ref{sec:phase_assignment}}.
During the DFF insertion step, optimal DFF insertion is determined for each datapath, as described in Section~\ref{sec:dff_insertion}. 

% With a single phase, this requirement reduces to full path balancing requirement. 
% any pair of clocked elements should be separated by at most $n$ 

% In single-phase clocking, $S(g)$ corresponds to the number of clocked elements traversed by a data signal before reaching $g$. 
% With MPC, however, the number of clocked gates traversed by a data signal to reach $g$ may be greater than $S(g)$, allowing more advanced processing to be performed with fewer clock stages.
% To ensure the correct order of data

% Observe that in single-phase clocking, the $\sigma(g)$ simply reduces to level of the gate.
% By separating the clocked elements, the data can be transferred across multiple phases
% , the number of path balancing DFFs can be reduced as compared to conventional synchronization. 

% The multiphase clocking has been effectively utilized in~\cite{li_beerel} to 

% Mention phase skipping in AQFP
% Mention degradation in throughput 

\section{Phase assignment}\label{sec:phase_assignment}

The phase assignment is the procedure of determining the stage (i.e., epoch and phase) of each gate within the network. 
An ILP-based methodology for phase assignment to reduce the number of path balancing DFFs is proposed in~\cite{li_beerel},
% The formulation in~\cite{li_beerel} can be represented as 
\begin{equation}\label{eq:beerel_cost_func}
\min\limits_{\sigma(g)\ \forall g\in \mG}\sum\limits_{(i,j)\in \mE} \left\lfloor \frac{\sigma(j)-\sigma(i)}{n} \right\rfloor,
\end{equation}
\begin{equation*}
\text{Subject to: }
\end{equation*}
\begin{equation}
\sigma(i) < \sigma(j) \hem\forall (i,j)\in \mE,
\end{equation}
\begin{equation}
\left\lfloor \frac{\sigma(i)}{n} \right\rfloor = 0  \hem\forall i\in \mI,
\end{equation}
\begin{equation}
\left\lfloor \frac{\sigma(i)}{n} \right\rfloor = \left\lfloor \frac{\sigma(j)}{n} \right\rfloor \hem\forall i,j\in \mO.
\end{equation}

This formulation is highly effective when considering the systems consisting of only clocked gates, i.e. $\mG=\mG_{\cfm{AS}}$, $\mG_{\cfm{AA}}=\mG_{\cfm{SA}}=\varnothing$
% Consider the example $2$-phase system depicted in Fig.~\ref{example_ILP_bad}a. 
% The number of DFFs is correctly determined using the cost function (\ref{eq:beerel_cost_func}).
This formulation however offers limited support for the unclocked gates, such as mergers and \cf{SA} \cf{AND} gate.
% misrepresent the required number of DFFs if the datapath consists of unclocked gates, as depicted in Fig.~\ref{example_ILP_bad}a. 
% % % % % % % % % % % % % % % % % % % % % % % % % % % % % % % % 
% Change example to merger gate
% % % % % % % % % % % % % % % % % % % % % % % % % % % % % % % % 
% Based on cost function (\ref{eq:beerel_cost_func}), the circuit requires two DFFs, since the presence of the merger is not considered in the model. 
% This datapath can however be balanced using only a single DFFs placed before the splitter, making both paths share the DFF.
Unclocked elements can greatly improve the quality of SFQ technology mapping by enriching the functionality available within a single clock cycle. 
In~\cite{bairamkulov_vlsisoc23}, for example, the area of the circuit was reduced by up to 54\% by utilizing asynchronous elements.
However, the unclocked gates impose timing constraints different from the clocked gates.
To extend the support of this ILP model to unclocked gates, additional constraints are introduced in this section.

\begin{figure}[b]
\centering
\includegraphics[width=\columnwidth]{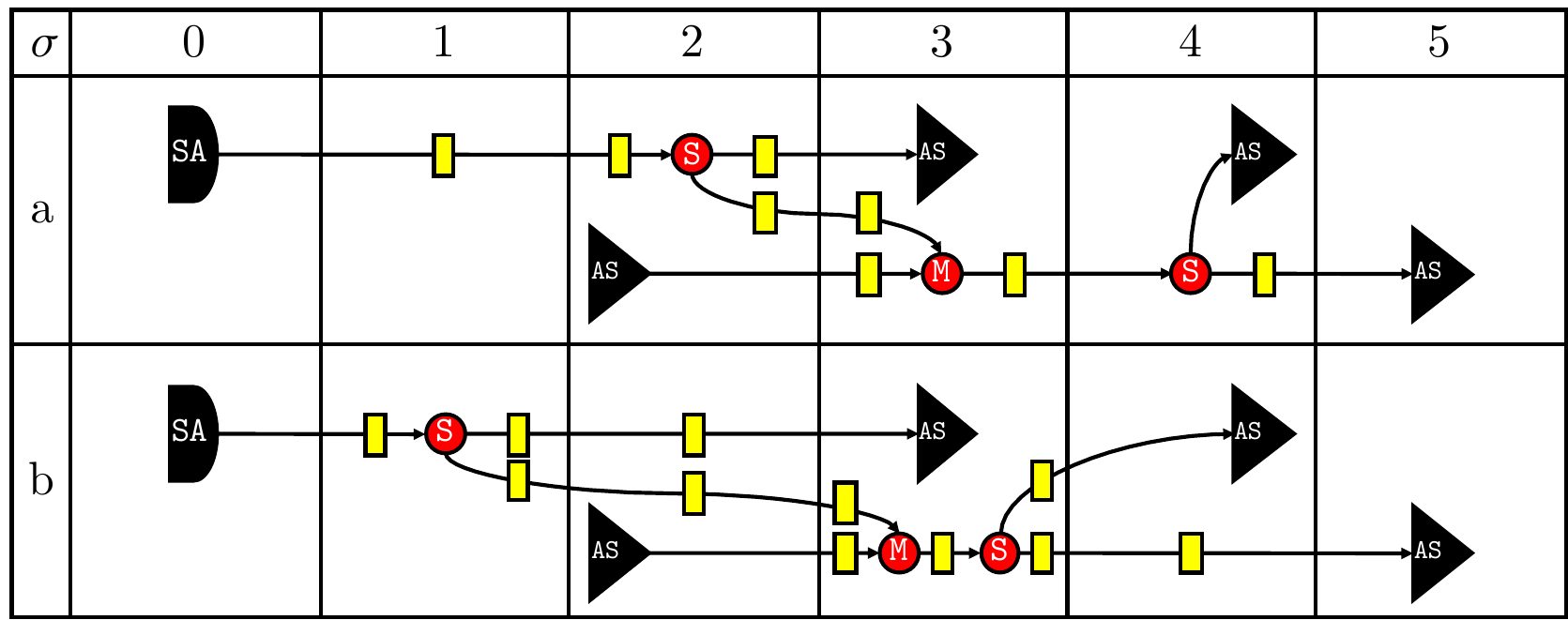}
\caption{
The effect of phase assigned to splitters and mergers. 
Datapaths (a) and (b) execute the same function.
The phases assigned to the splitters and mergers (denoted as $\mathtt{S}$ and $\mathtt{M}$, respectively) are however different.
The potential DFF locations (depicted as yellow rectangles) are therefore different in the two datapaths.
}
\label{fig:splitter_merger}
\end{figure}

While the notions of phase, epoch, and stage are primarily relevant for the clocked gates, 
% the unclocked gates are also assigned a stage.
it is however important to also assign a stage to the \cf{AA} and \cf{SA} elements, since a stage of an unclocked element guides the DFF insertion.
Consider two networks containing an asynchronous datapath illustrated in Fig.~\ref{fig:splitter_merger}.
Both networks realize the same function, however, the possible locations of path balancing DFFs are different. 
Assigning a stage to an unclocked gate is important to correctly place the path balancing DFFs.
% , the location of path balancing DFFs within the circuit can be correctly considered. 

% Recall that the SFQ logic gates can be divided into three categories, namely, \emph{\cf{AA}}, \emph{\cf{AS}}, and \emph{\cf{SA}}, where 
% the first letter denotes whether input signals should arrive (a)synchronously, while the second letter indicates whether the output is released (a)synchronously.
% Each category imposes different timing constraints on data propagation.

\subsection{\cf{SA} element constraints}

\begin{figure}
\centering
\stackunder{\includegraphics[scale=0.35]{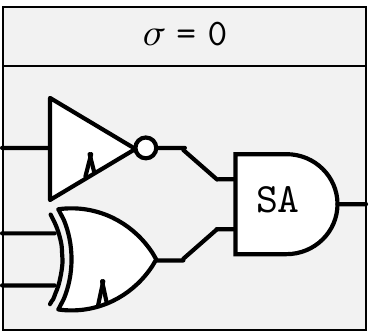}}{a)}\hfill
\stackunder{\includegraphics[scale=0.35]{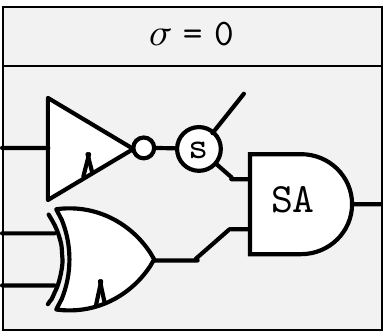}}{b)}\hfill
\stackunder{\includegraphics[scale=0.35]{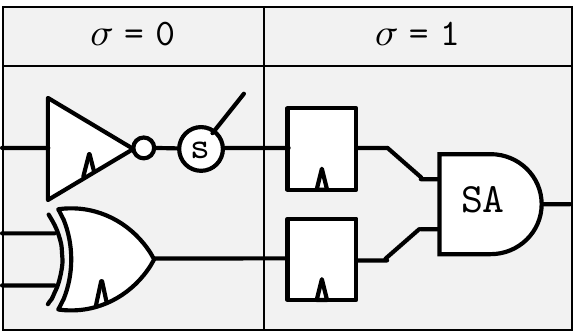}}{c)}
\caption{
Timing constraints of an \cf{SA} gate.
a) The gate should be preceded by an \cf{AS} gate at the same stage. 
b) Example of an invalid assignment. 
The \cf{AS} gate is preceded by an \cf{AA} gate at the same stage.
c) The issue is resolved by shifting the \cf{SA} gate to the next stage and inserting two path balancing DFFs.
}
\label{fig:sa_constr}
\end{figure}

Recall that the inputs of the \cf{SA} gates require simultaneous arrival of data pulses to the gate. 
To ensure the simultaneous arrival, the \cf{AS} gate should be placed immediately before the \cf{SA} gate, at the same clock stage, as shown in Fig.~\ref{fig:sa_constr}a.
% Since the \cf{SA} gate is not clocked, the \cf{SA} gate can be placed at the same stage as its fanin \cf{AS} gate. 
If the stage of \cf{AS} gate is earlier than the \cf{SA} gate, a path balancing DFF should be inserted. 
In addition, the \cf{SA} gate cannot be placed at the same stage as the fanin, if the fanin is not an \cf{AS} gate. 
% This additional DFF allows the \cf{SA} gate to be considered as \cf{AS} gate.
For example, if the fanin of the \cf{SA} gate has fanout greater than one, a splitter is inserted after the \cf{AS} gate.
Thus, the fanin type of the \cf{SA} gate is a splitter (\cf{AA} gate). 
The propagation delay incurred by the \cf{AA} gates, such as splitters, may desynchronize the inputs arriving to the \cf{SA} gate \cite{bairamkulov_vlsisoc23}, producing a data hazard, as illustrated in Fig.~\ref{fig:sa_constr}b.
To avoid this condition, the minimum stage of gate $g$ is increased by one, as shown in Fig.~\ref{fig:sa_constr}c.

\subsection{Asynchronous paths}

The sharing of DFFs by a complex unclocked path has been highlighed as early as 1991 in~\cite{leiserson_1991}, where sharing of DFFs is maximized to minimize the circuit area.
% , as illustrated in Fig.~\ref{fig:shared_dff}a
% A similar formulation is possible to estimate the number DFFs required for a path with multiple fanout
% Due to the DFF sharing in a path with multiple fanouts, the number of DFFs is determined by the fanout with the latest stage,
% Basically, need to say that estimating the number precisely is difficult
While estimating the minimum number of DFFs in special cases, such as splitter tree, is relatively simple, more complex asynchronous paths, however, require more sophisticated processing and, hence, prohibitive runtime.
% As described in section~\ref{SAT_BASED_DFF_PLACEMENT}, precisely estimating the number of DFFs required by complex asynchronous path requires  processing and prohibitive runtime. 
During the phase assignment stage we therefore approximate the number of DFFs as
\begin{equation}
    C \left( \sigma \left( g \right) \right) = \left\lfloor \frac{\max\limits_{a\in FO \left(g \right)} \left( \sigma \left( a \right) \right) - \sigma \left( g \right)}{n} \right\rfloor.
    % \sigma \left( a \right)+ \left( a\notin \mG_{\cf{AS}} \right)\hem\forall a\in FI \left( g \right),
\end{equation}
The exact number of DFFs is determined during the DFF insertion stage described in Section~\ref{sec:dff_insertion}. 
% After considering the influence of the unclocked gates on the objective and constraint functions, t
The combined optimization problem is formulated as 
% After considering the influence of the unclocked gates to the objective and constraint functions, the combined optimization problem is formulated as 
\begin{equation}\label{eq:combined_cost_func}
\min\limits_{\sigma \left( g \right)\hem\forall g\in \mG}
\sum\limits_{ \left( i,j \right)\in \mE} \left\lfloor \frac{\sigma \left( j \right)-\sigma \left( i \right) + \left( j\in \mG_{\cfm{SA}} \right)}{n} \right\rfloor,
\end{equation}
\begin{equation*}
\text{Subject to: }
\end{equation*}
\begin{equation}
\sigma \left( i \right) <    \sigma \left( j \right) \hem\forall \left( i,j \right)\in \mE, j\in \mG_{\cfm{AS}},
\end{equation}
\begin{equation}
\sigma \left( i \right) \leq \sigma \left( j \right) \hem\forall \left( i,j \right)\in \mE, j\notin \mG_{\cfm{AS}},
\end{equation}
\begin{equation}
    \sigma \left( g \right) \geq \sigma \left( a \right)+ \left( a\notin \mG_{\cfm{AS}} \right)\hem\forall g\in \mG_{\cfm{SA}}, a\in FI \left( g \right),
\end{equation}
% \begin{equation}
%     \sigma \left( g \right) \geq \sigma \left( a \right)+ \left( a\notin \mG_{\cfm{AS}} \right)\hem\forall g\in \mG_{\cfm{SA}}, a\in FI \left( g \right),
% \end{equation}
\begin{equation}
\left\lfloor \frac{\sigma \left( i \right)}{n} \right\rfloor = 0  \hem\forall i\in \mI,
\end{equation}
\begin{equation}
\left\lfloor \frac{\sigma \left( i \right)}{n} \right\rfloor = \left\lfloor \frac{\sigma \left( j \right)}{n} \right\rfloor  \hem\forall i,j\in \mO.
\end{equation}

Each gate $g$ with $\#FO(g) > 1$ requires a splitter tree to copy the signal to each of the fanouts. 
After completing the phase assignment, a splitter tree that maximizes sharing of path balancing DFFs can be produced. 
%  can be efficiently placed after each multifanout gate.
First, the fanouts $FO(g)$ of a gate $g$ are sorted by phase in the ascending order, producing a sequence $A=[a_0, a_1,\cdots,a_k]$, \mbox{$\sigma(a_i) \leq \sigma(a_j) \iff i<j$}.
Next, the splitters $[s_0,...,s_{k-1}]$ are inserted at phases $a_0,\cdots,a_{k-1}$. 
% The splitter $s_0$ has a gate $g$ as a fanin.
Each splitter $s_i$ has a splitter $s_{i-1}$ as a fanin, except $s_0$ whose fanin is the gate $g$. 
Each splitter $s_i$ has the $a_i$ and $s_{i+1}$ as a fanout, except $s_{k-1}$ whose fanouts are $a_{k-1}$ and $a_k$.
Using this method, the splitters are placed as late as possible within the network, maximizing sharing of the datapath.

\section{DFF placement}\label{sec:dff_insertion}

The complex asynchronous datapaths greatly complicate the DFF insertion process. 
Consider for example the datapath shown in Fig.~\ref{fig:splitter_merger}. 
The optimal placement of the DFFs cannot be easily inferred from the datapath topology. 
In this section, we propose a CP-SAT based placement methodology for determining the optimal location of the DFFs.
Our methodology consists of two stages.
First, we identify \textit{independent datapaths} within an unbalanced logic network, as described in Subsection~\ref{subsec:indep_path}.
Next, for each independent datapaths, we formulate a CP-SAT problem where the timing constraints are satisfied using the minimum number of DFFs, as described in Subsection~\ref{subsec:cp_sat}.

\subsection{Independent datapaths}\label{subsec:indep_path}

Recall that only a single clocked gate can be placed at the same stage along a datapath. 
% Observe that the \cf{AS} and \cf{SA} elements strictly determine the placement of clocked elements along a datapath, as shown in Fig.~\ref{fig:datapath_AS_SA}.
The \cf{AS} and \cf{SA} elements therefore prevent a DFF from being placed at the same stage. 
% Furthermore, this certainty in DFFs placement effectively prevents the DFF placement before and after an \cf{AS}/\cf{SA} element, from affecting each other.
% IMPORTANT : REPHRASE IT'S HORRIBLE!
Furthermore, any timing constraint imposed by a DFF placed before the \cf{AS}/\cf{SA} element, will be dominated by the timing constraint imposed by the \cf{AS}/\cf{SA} element itself at any location after the \cf{AS}/\cf{SA} element.
This feature allows us to represent a network-wide DFF insertion problem as multiple smaller problems.
This partitioning capability brings two major advantages to the DFF insertion process. 
First, while the network-wide DFF insertion problem can be solved in $O(e^{N})$ time, the complexity of a partitioned problem is $O(e^{k}N_{path})$, where $N$ is the network size, $N_{path}\propto N$ is the number of independent datapaths and $k$ is the size of the largest partition (typically, $k\ll N$).
Second, partitioning facilitates multithreaded processing, allowing faster processing on a multicore system.

The proposed DFF insertion method is applied to a single independent datapath constrained by the synchronous gates at the input and the output, as illustrated in Fig.~\ref{fig:datapath_example}a. 
The datapath $P$ can be described as a portion of the logic network $P=(I,A,O)$, where set $A\subseteq \mG_{\cfm{AA}}$ contains the \cf{AA} gates within the datapath and  sets $I,O\subseteq \mG_{\cfm{AS}}\cup \mG_{\cfm{SA}}$ describe, respectively, the \cf{AS} and \cf{SA} gates at the input and output of the datapath. 
In Fig.~\ref{fig:datapath_example}a, the three independent paths are 
\begin{equation*}
    P_1=\left(I=\{\mathtt{A}, \mathtt{B}, \mathtt{D}\}, A=\{\binone, \mathtt{2}, \mathtt{3}, \mathtt{4}, \mathtt{5}\}, O=\{\mathtt{W}, \mathtt{X}, \mathtt{Y}, \mathtt{Z}\}\right),
\end{equation*}
\begin{equation*}
    P_2=\left(I=\{\mathtt{E}\}, A=\varnothing, O=\{\mathtt{W}\}\right),
\end{equation*}
\begin{equation*}
    P_3=\left(I=\{\mathtt{C}, \mathtt{F}\}, A=\{\mathtt{6}\}, O=\{\mathtt{Z}\}\right).
\end{equation*}

\subsection{DFF insertion}\label{subsec:cp_sat}

\begin{figure}
\centering
% \stackunder[1pt]{
% \includegraphics[width=0.484\textwidth]{Figures/datapath_1_a.pdf}
% }{a)}
% \stackunder[1pt]{
% \includegraphics[width=0.501\textwidth]{Figures/datapath_2_b.pdf}
% }{b)}
\includegraphics[width=\columnwidth]{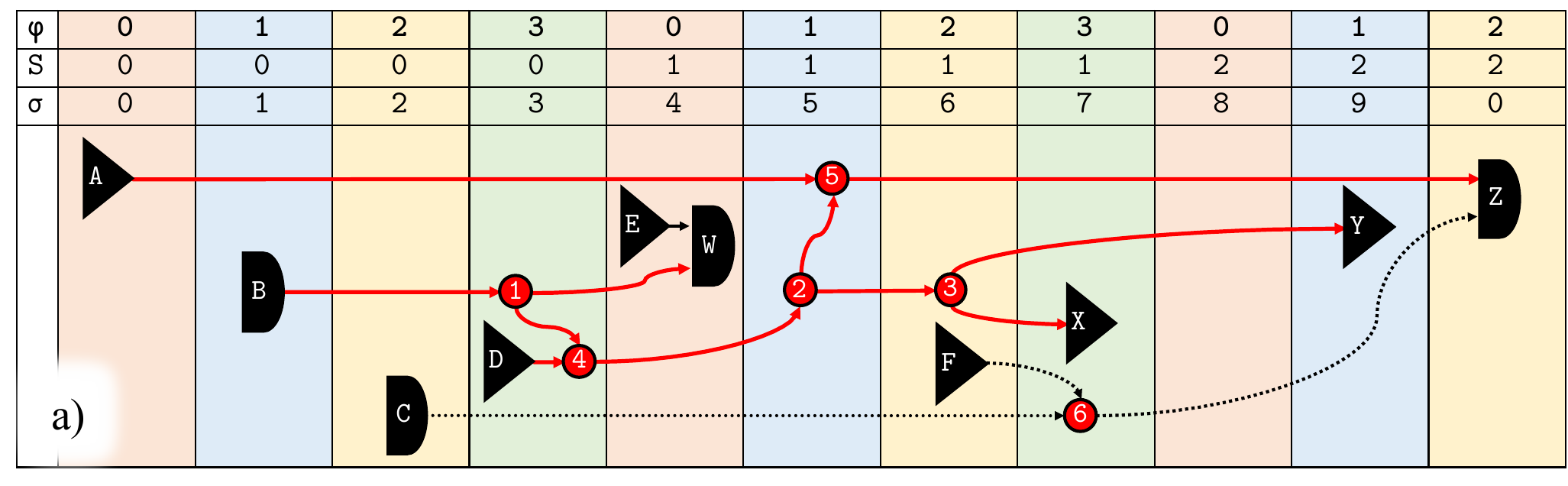}\\
\includegraphics[width=\columnwidth]{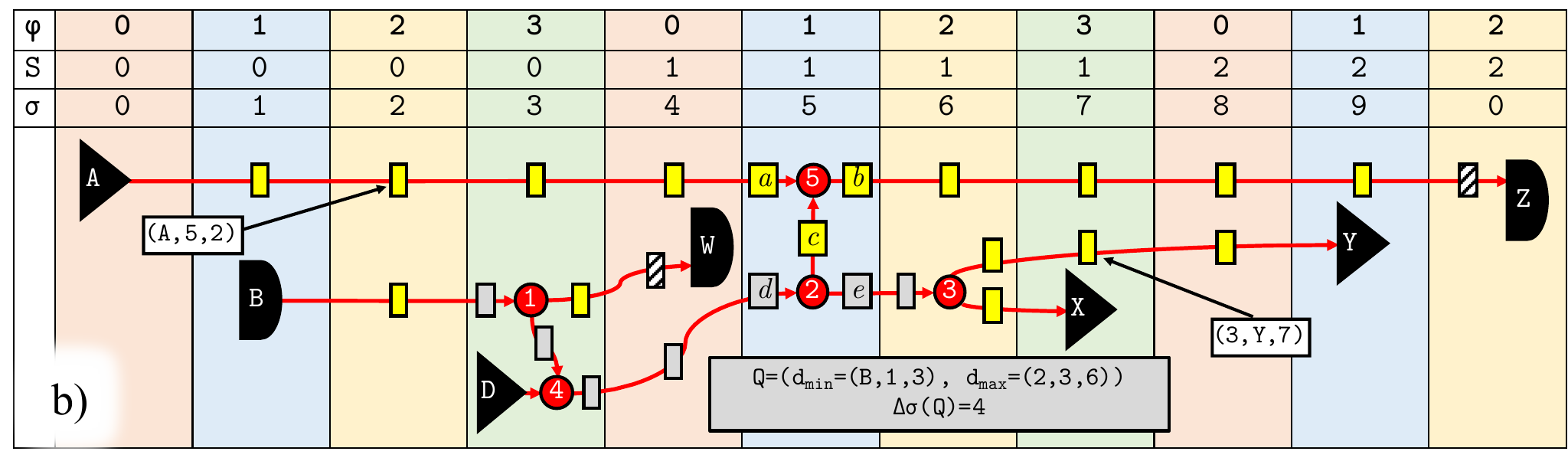}\\
\vspace{-1em}
\caption{Example of an asynchronous datapath within a four-phase network ($n=4$).
The black triangles, and black curved shapes represent the \cf{AS} and \cf{SA} elements, respectively.
Red circles denote the \cf{AA} elements.
a) Three independent paths are drawn with red solid, black solid, and black dotted arrows.
b) DFF sites shown as rectangles within the red independent path. 
The grey DFF sites represent the chain $\mathtt{Q}$ of length $n$. 
Diagonally shaded DFF sites precede the \cf{SA} gates. 
DFFs should therefore be placed at these sites.
}
% \hline
% \vspace{-1.5em}
\label{fig:datapath_example}
\end{figure}

After identifying each independent path within the network, we determine the potential DFF sites for subsequent DFF insertion. 
% Observe that the asynchronous elements produce an ambiguity in DFF placement. 
To uniquely identify each potential DFF site, we define $d=(g_d^i, g_d^o, \sigma(d))$, where $g_d^i\in I\cup A$ and  $g_d^o\in A\cup O$ are the fanin and fanout elements of the DFFs and $\sigma(d)$ is the stage of the DFF site.
We define a chain $Q=(d_{min}, \cdots, d_{max})$ as a sequence of sequentially adjacent DFF locations situated between $d_{min}$ and $d_{max}$.
The examples of DFF sites and a chain are shown in Fig~\ref{fig:datapath_example}b. 
For convenience, we define the length of chain $\Delta\sigma(Q)$ as the stage difference between $d_{min}$ and $d_{max}$.
% The phases $\sigma_{min}(Q)$ and $\sigma_{max}(Q)$ denote the stage of $d_{min}$ and $d_{max}$ within the chain $Q$.
% Not that any reconvergent path produces multiple chains with the same starting . 
% The DFF locations and the corresponding chains within the datapath are shown in Fig.~\ref{fig:datapath_example}. 

For each DFF site, we introduce a binary variable $\delta(d)$ equal to $\binone$ if the DFF is placed at $d$ and $\binzero$ otherwise.
The problem of minimizing the number of path balancing DFFs can therefore be formulated as a CP-SAT problem minimizing
\begin{equation}\label{eq:condition_1}
    \sum\limits_{d\in P} \delta(d).
\end{equation}
Several constraints describe the valid placement of the DFFs within an independent path.
First, for each stage, only a single DFF can be placed along a chain. 
% , as illustrated in Fig.~\ref{fig:datapath_example}c
In Fig.~\ref{fig:datapath_example}b, for example, the stage $\sigma=5$ has five potential DFF locations, $a,b,c,d$, and $e$. 
The DFFs here should however satisfy the constraint
\begin{equation}\label{eq:condition_2}
\sum\limits_{d\in Q, \sigma_d=i} \delta(d) \leq 1.
\end{equation}
The combinations of DFF sites satisfying this constraint are $\{a,c,e\}$, $\{a,d\}$, $\{b,e\}$, since only these sets do not place multiple DFFs along any chain.
Other conflicting DFF sites can be found at stages 3 and 6.
% are highlighted in blue in Fig.~\ref{fig:datapath_example}b.

Second, the distance between any pair of clocked elements along a chain should not exceed $n$ in an $n$-phase system.
This requirement can be formulated as a condition
\begin{equation}\label{eq:condition_3}
    \bigvee\limits_{d\in Q, \Delta\sigma(Q)=n}  \delta(d) = \binone
\end{equation}
stating that at least one DFF should be placed along any chain of length $n$.

Finally, recall that the \cf{SA} gates should be preceded by an \cf{AS} gate to function correctly. 
% Therefore, the DFFs should be placed before the \cf{SA} gates
% Thus, the \cf{SA} elements require the \cf{AS} gate to be placed directly before the inputs.
Thus, if an \cf{SA} element is not directly preceded by an \cf{AS} gate, a DFF should be placed before the \cf{SA} element, 
\begin{equation}\label{eq:condition_4}
    \delta(d) = \binone \hem\forall g_d^o\in \mG_{\cfm{SA}}, \sigma(g_d)=\sigma_d.
\end{equation}
The DFF sites before the \cf{SA} gates are shaded with diagonal hatch pattern in Fig.~\ref{fig:datapath_example}b. 

Equations (\ref{eq:condition_1})-(\ref{eq:condition_4}) constitute a CP-SAT problem where the conditions (\ref{eq:condition_2})-(\ref{eq:condition_4}) are satisfied with the minimum number of DFFs (\ref{eq:condition_1}).
% where $\sigma(g_d)$ denotes the phase of the gate $g_d$.
% Consider, for example, the \cf{AND} gate $c$ in Fig.~\ref{fig:datapath_example}.
% The first fanin of the gate is directly preceded by an \cf{AS} gate (\cf{XOR}).
% Therefore, no DFF is needed nor can be placed at this location. 
% The second fanin of the gate however requires a path balancing DFF to be placed before the \cf{SA} gate. 

\section{Experimental results}\label{sec:validation}

We integrate the DFF placement methodology into the technology mapping flow for SFQ compound gate circuits. 
The SFQ circuits are synthesized with \cf{mockturtle} using a database of precomputed compound gate structures~\cite{bairamkulov_vlsisoc23}. 
The circuits are next decomposed into the functional primitives while removing all DFFs within the circuit. 
We next apply our CP-SAT-based phase assignment and DFF insertion procedures using Google OR-Tools~\cite{ortools}.
% Each CP-SAT problem run has been limited by 60 seconds to ensure no excessive runtime.

We apply our mapping flow to synthesize a subset of EPFL~\cite{soeken2018epfl} and ISCAS~\cite{hansen1999unveiling} benchmark circuits.
For each benchmark circuit, the total runtime does not exceed 60.64 seconds. 
Phase assignment is a major component of the total runtime.
Optimal phase assignment has been successfully determined in all circuits.
In most cases, the phase assignment is completed in under 5 seconds, with the longest runtime (42 seconds) required by the \cf{voter} circuit.
% The solution to the DFF insertion specifies the location and phase of each DFF inserted into the network.
Naturally, larger circuits, such as \cf{s13207}, \cf{voter} and \cf{priority} require longer runtime for phase assignment. 
Applying our tool to larger circuits would require additional processing time.
% (e.g., \cf{hyp} from EPFL benchmarks)
Note that although finding the optimal phase assignment may require prohibitive runtime, a feasible solution is often available much earlier.
% Furthermore, a suboptimal phase assignment has no direct effect on the DFF insertion process. 
Therefore, by relaxing the optimality requirement, the our framework can support larger circuits.
\footnote{Mapping of \cf{hyp} (the largest circuit in EPFL suite) has been completed within one hour with only 15 and 20 minutes allocated to phase assignment and DFF insertion, respectively 
(not presented due to space constraints).}

\begin{table*}[h]
\setlength\tabcolsep{4pt}
\tabcolsep=2.5pt
\centering
\caption{Comparison of multiphase clocking with dual clocking method \cite{dual_clock} for different throughputs}
\vspace{-1em}
\begin{tabular}{l|rr|rr|rr|r|rr|rr|rr|r}
\toprule 
               & \mcl{7}{c|}{1/7 throughput}                                                              & \mcl{7}{c}{1/12 throughput}                                                                 \\
\cmidrule(rl){2-8} \cmidrule(rl){9-15}
% \cmidrule(rl){2-3} \cmidrule(rl){4-5} \cmidrule(rl){6-7} \cmidrule(rl){8-8} \cmidrule(rl){9-10} \cmidrule(rl){11-12} \cmidrule(rl){13-14} \cmidrule(rl){15-15}
               & \mcl{2}{c|}{DCM}                 & \mcl{2}{c|}{Multiphase}           & \mcl{2}{c|}{Change} & Runtime & \mcl{2}{c}{DCM}                    & \mcl{2}{c}{Multiphase}          & \mcl{2}{c}{Change}  & Runtime  \\
               % \rowcolor{gray!15}
               & \#DFF  & \#JJ                   & \#DFF   & \#JJ                   & \#DFF    & \#JJ     & (s)     & \#DFF     & \#JJ                   & \#DFF  & \#JJ                   & \#DFF     & \#JJ    & (s)     \\
\cmidrule(rl){2-3} \cmidrule(rl){4-5} \cmidrule(rl){6-7} \cmidrule(rl){8-8} \cmidrule(rl){9-10} \cmidrule(rl){11-12} \cmidrule(rl){13-14} \cmidrule(rl){15-15}
\cf{int2float} & 117    & \mcl{1}{r|}{7'770}     & 217     & \mcl{1}{r|}{5'136}     & +85\% &\mcl{1}{r|}{--34\%} & 4.42 & 39        & \mcl{1}{r|}{5'140}     & 46     & \mcl{1}{r|}{3'939}     & +18\%  &\mcl{1}{r|}{--23\%} & 4.13 \\
\rowcolor{gray!15}
\cf{priority}  & 8'562  & \mcl{1}{r|}{257'252}   & 3'285   & \mcl{1}{r|}{45'094}    &--62\% &\mcl{1}{r|}{--82\%} & 54.01 & 4'225     & \mcl{1}{r|}{158'568}   & 1'775  & \mcl{1}{r|}{34'524}    &--58\%  &\mcl{1}{r|}{--78\%} & 30.33 \\
\cf{voter}     & 7'204  & \mcl{1}{r|}{447'044}   & 2'180   & \mcl{1}{r|}{162'804}   &--70\% &\mcl{1}{r|}{--64\%} & 60.64 & 3'732     & \mcl{1}{r|}{355'144}   & 1'568  & \mcl{1}{r|}{158'520}   &--58\%  &\mcl{1}{r|}{--55\%} & 48.94 \\
\rowcolor{gray!15}
\cf{c432}      & 224    & \mcl{1}{r|}{10'734}    & 342     & \mcl{1}{r|}{5'116}     & +53\% &\mcl{1}{r|}{--52\%} & 14.92 & 118       & \mcl{1}{r|}{7'124}     & 240    & \mcl{1}{r|}{4'402}     & +103\% &\mcl{1}{r|}{--38\%} & 10.14 \\
\cf{c880}      & 362    & \mcl{1}{r|}{14'658}    & 254     & \mcl{1}{r|}{6'190}     &--30\% &\mcl{1}{r|}{--58\%} & 7.56 & 187       & \mcl{1}{r|}{9'483}     & 119    & \mcl{1}{r|}{5'245}     &--36\%  &\mcl{1}{r|}{--45\%} & 7.25 \\
\rowcolor{gray!15}
\cf{c1908}     & 282    & \mcl{1}{r|}{13'169}    & 125     & \mcl{1}{r|}{3'529}     &--56\% &\mcl{1}{r|}{--73\%} & 10.22 & 144       & \mcl{1}{r|}{8'739}     & 69     & \mcl{1}{r|}{3'137}     &--52\%  &\mcl{1}{r|}{--64\%} & 8.78 \\
\cf{c3540}     & 776    & \mcl{1}{r|}{43'437}    & 589     & \mcl{1}{r|}{17'016}    &--24\% &\mcl{1}{r|}{--61\%} & 11.83 & 282       & \mcl{1}{r|}{26'897}    & 440    & \mcl{1}{r|}{15'973}    & +56\%  &\mcl{1}{r|}{--41\%} & 9.29 \\
\rowcolor{gray!15}
\cf{c1355}     & 193    & \mcl{1}{r|}{8'739}     & 46      & \mcl{1}{r|}{4'515}     &--76\% &\mcl{1}{r|}{--48\%} & 4.18 & 119       & \mcl{1}{r|}{6'149}     & 44     & \mcl{1}{r|}{4'501}     &--63\%  &\mcl{1}{r|}{--27\%} & 3.89 \\
\cf{s13207}    & 1'795  & \mcl{1}{r|}{106'346}   & 1'837   & \mcl{1}{r|}{42'382}    & +2\%  &\mcl{1}{r|}{--60\%} & 27.87 & 571       & \mcl{1}{r|}{60'766}    & 1'082  & \mcl{1}{r|}{37'097}    & +89\%  &\mcl{1}{r|}{--39\%} & 25.90 \\
\rowcolor{gray!15}
\cf{s5378}     & 645    & \mcl{1}{r|}{50'766}    & 808     & \mcl{1}{r|}{21'761}    & +25\% &\mcl{1}{r|}{--57\%} & 7.87 & 255       & \mcl{1}{r|}{34'053}    & 368    & \mcl{1}{r|}{18'681}    & +44\%  &\mcl{1}{r|}{--45\%} & 7.07 \\
\cf{s382}      & 56     & \mcl{1}{r|}{4'448}     & 89      & \mcl{1}{r|}{2'411}     & +59\% &\mcl{1}{r|}{--46\%} & 4.17 & 9         & \mcl{1}{r|}{2'750}     & 48     & \mcl{1}{r|}{2'124}     & +433\% &\mcl{1}{r|}{--23\%} & 3.93 \\
\cmidrule(rl){1-1} \cmidrule(rl){2-8} \cmidrule(rl){9-15}
Geomean        & 557.04 & \mcl{1}{r|}{29'868.20} & 413.48  & \mcl{1}{r|}{11'965.54} &--25.8\% &\mcl{1}{r|}{--59.9\%} & 12.02 & 227.85    & \mcl{1}{r|}{19'565.38} & 229.57 & \mcl{1}{r|}{10'467.10} & +0.8\%   &\mcl{1}{r|}{--47.5\%} & 10.03 \\
\bottomrule
\end{tabular}\label{tab:with_dcm}
\vspace{-1em}
\end{table*}

\begin{figure*}
\centering
\includegraphics[height=13em]{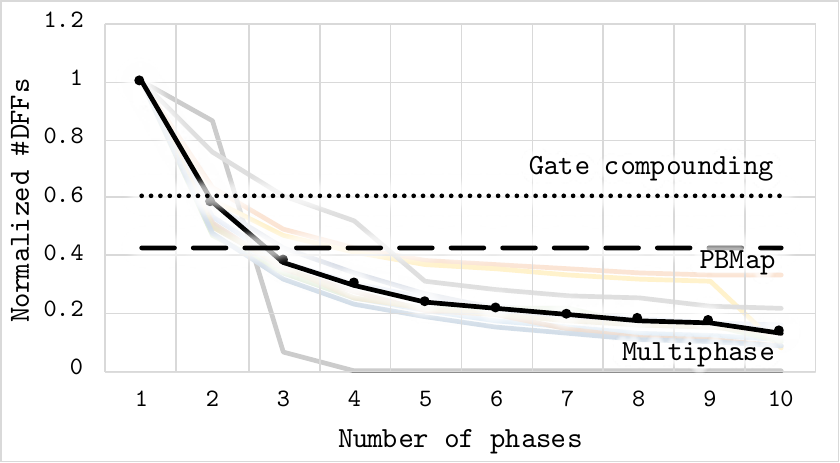} \hfill
\includegraphics[height=13em]{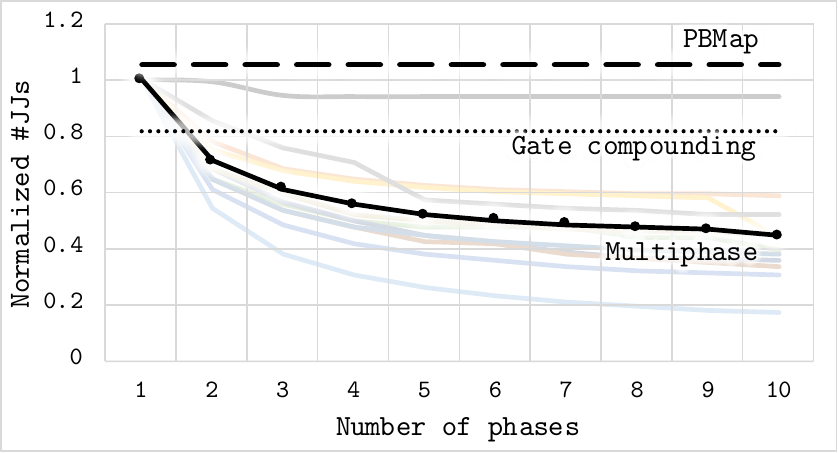}
\vspace{-0.5em}
\caption{
Normalized average number of DFFs (left) and JJs (right) with different number of phases.
% On average, three phases are sufficient to outperform both gate compounding and PBMap.
The lines in the background represent normalized number of DFFs and JJs for individual benchmarks. 
}
\vspace{.3em}
\hrule
\vspace{-1.5em}
\label{fig:ndff_njj_comparison}
\end{figure*}

% The results are shown in Table.~\ref{}.
We first compare our results with dual clocking method~\cite{pasandi_acm}.
Two experiments are reported where the frequency of the slow clock is 7 and 12 times smaller than the frequency of the fast clock. 
% In both cases,  as compared to full path balancing. 
For fair comparison, we use 7 and 12 phases to equalize the throughput.
The results are shown in Table~\ref{tab:with_dcm}.
Note that unlike~\cite{pasandi_acm}, the DFFs within the \cf{AND} and \cf{OR} gate are explicitly counted in our work due to the use of the unclocked (\cf{SA}) \cf{AND} and \cf{OR} elements.
Furthermore, additional elements, such as pulse repeaters and NDRO flip-flops are needed in DCM.
% The number of extra DFFs is reduced by an average of 19.2\%, not considering the NDRO DFFs required by the dual clocking method.
We therefore focus on the number of JJs, commonly used in SFQ technology as an area metric. 
The number of JJs is reduced, on average, by 59.9\% when using 7-phase clocking.
The largest reduction is observed in the \cf{priority} benchmark circuit containing many imbalanced paths that require long chains of path balancing DFFs to be inserted. 
With multiphase clocking, these chains can be balanced by approximately 7 times fewer DFFs, contributing to the area savings.

For 12-phase clocking, the area is reduced by 47.5\%. 
Observe that the smallest reduction in area (and the increase in DFF count) are observed in circuits with relatively small depth.
These circuits therefore have fewer imbalanced paths that could benefit from multiphase clocking.
Despite relatively large number of DFFs, we still manage to reduce the total JJ count due to the use of unclocked elements (\cf{AA} and \cf{SA}) supported by our tool. 

% DRO DFF count is increased in primarily  \cf{int2float}, \cf{c432} and \cf{c3540}.
% This poor performance of multiphase clocking can be attributed to relatively small depth of these circuits and, hence, fewer imbalanced paths that could benefit from multiphase clocking.

The Fig.~\ref{fig:ndff_njj_comparison} (left) describes the relationship between the number of DFFs and the number of phases.
The numbers of DFFs are normalized with respect to a single phase system mapped with our flow. 
Consistent with~\cite{li_beerel} the number of DFFs reduces with multiple phases.
The area savings due to an additional phase gradually diminishes as more phases are added. 
We compare our results against PBMap~\cite{pbmap2019} and gate compounding~\cite{bairamkulov_vlsisoc23}, the state-of-the-art single-phase SFQ mapping algorithms.
With our mapping flow, two phases are sufficient, on average, to outperform gate compounding.
With three phases our flow yields fewer DFFs than PBMap. 
The Fig.~\ref{fig:ndff_njj_comparison} (right) describes the relationship between the circuit area (expressed as the number of JJs) and the number of phases.
Our flow outperforms PBMap with a single phase, primarily due to the use of asynchronous gates (e.g., merger instead of an \cf{OR} gate).
A two-phase network synthesized with our flow requires fewer JJs than a single-phase compound-gate network.

% \begin{figure}
% \centering
% \includegraphics[width=0.9\columnwidth]{Figures/mph_avg_ndff.pdf}
% \caption{
% Normalized average number of DFFs with different number of phases.
% On average, three phases are sufficient to outperform both gate compounding and PBMap.
% The lines in the background represent normalized number of DFFs for individual benchmarks. }
% \label{fig:ndff_comparison}
% \end{figure}

% In RSFQ, the circuit area is commonly expressed as the number of JJs within the network.

% \begin{figure}
% \centering
% \includegraphics[width=0.9\columnwidth]{Figures/mph_avg_area.pdf}
% \caption{
% Normalized average number of JJs (area metric) with different number of phases.
% On average, two phases are sufficient to outperform both gate compounding and PBMap.
% The lines in the background represent normalized number of JJs for individual benchmarks. }
% \label{fig:njj_comparison}
% \end{figure}

\section{Conclusions}\label{sec:conclusions}

RSFQ technology presents a remarkable opportunity to achieve unprecedented performance and energy efficiency of computing systems.
Realizing its full potential however requires overcoming major technological issues, including path balancing.
Multiphase clocking can substantially reduce the path balancing overhead in SFQ systems.
Existing path balancing tools however offer limited support of multiphase clocking in systems with asynchronous SFQ gates. 
In this work, we presented a technology mapping and path balancing methodology for multiphase SFQ systems based on constraint programming with satisfiability (CP-SAT). 
An SFQ logic network is initially synthesized with \cf{mockturtle} logic synthesis library. 
Next, the circuit is decomposed into primitives and the DFFs are removed from the network.
Each primitive gate is assigned a phase using the CP-SAT while satisfying the special timing constraints imposed by the unclocked SFQ elements.
Finally, using the novel CP-SAT formulation, the minimum number of path balancing DFFs satisfying the timing constraints is determined. 
In the experimental results, we showed an average of 59\% reduction in the number of JJs when compared to dual clocking method.
Furthermore, with only two phases, our methodology yields smaller networks than the state-of-the-art single-phase SFQ mapping techniques.

\scriptsize\bibliographystyle{sty/myIEEE.bst}
% \bibliography{bibliography.bib}

\end{document}